\documentclass[12pt]{article}

\input epsf
\begin {document}
\begin{titlepage}
\begin{flushright}
{\small
NYU-TH-05/09/3}
\end{flushright}
\vskip 0.9cm

\centerline{\Large \bf Cosmic Superstrings Stabilized by Fermions}
\vspace{.2cm}
\vspace{0.5cm}
\centerline{\Large \bf 
 }

\vskip 0.7cm
\centerline{\large Jose J. Blanco-Pillado\footnote{E-mail: 
blanco-pillado@physics.nyu.edu} 
and Alberto Iglesias\footnote{E-mail: iglesias@physics.nyu.edu}}
\vskip 0.3cm
\centerline{\em Center for Cosmology and Particle Physics}
\centerline{\em Department of Physics, New York University, New York, 
NY, 10003, USA}                                                          

\vskip 1.9cm 

\begin{abstract}

We show that there exist massive perturbative states of the ten dimensional
Green-Schwarz closed superstring that are stabilized against collapse
due to presence of fermionic zero modes on its worldsheet. The excited 
fermionic degrees of freedom backreact on the spacetime motion of the 
string in the same way as a neutral persistent current would, rendering 
these string loops stable. We point out that the existence of these states
could have important consequences as stable loops of cosmological size 
as well as long lived states within perturbative string theory.

\end{abstract}

\end{titlepage}

\newpage

\section{Introduction}

Despite their common origin, the study of fundamental strings and 
cosmic strings have been traditionally quite differentiated 
\cite{GSW-book,Polchinski-book,Alex-book}. On the other hand, 
this distinction may just turn out to be an artifact due to historical reasons. 
Early estimates of the possible role of fundamental strings as cosmic strings
were not too optimistic \cite{Witten-cosmicsuperstrings}, but recent developments 
in string theory models of the early universe seem to hint to the possibility 
that there could exist fundamental strings of cosmological 
size \cite{Tye-1,Tye-2,Gia-Alex-1,Gia-Alex-2,CMP}.
These fundamental strings would naturally behave as classical objects
rendering them effectively a new type of cosmic strings.

This new connection has been exploited in the literature to suggest
that we should actually look for signatures of string theory in the
cosmological observations that we are able to perform today. This
is undoubtedly a very interesting idea that clearly deserves a detailed
study. Most of the work in this direction has focused on the study 
of the characteristics of these new type of one-dimensional objects, trying to 
extract distinctive signatures of the network of strings that can
conclusively show that we are indeed dealing with string theory
models of the early universe.

In this letter we would like to explore the opposite route. We will
show that there exist quantum mechanical states of perturbative
string theory that mimic solutions of the equations
of motion of well known models of classical cosmic strings, namely 
superconducting cosmic strings \cite{Witten}. 

The paper is organized as follows. In Section 2, we introduce the
Green-Schwarz formalism for the superstring, mainly following the notation
of reference \cite{GSW-book} \footnote{The reader familiar with this 
subject may want to skip ahead to Section 3.}. In Section 3 we discuss
the simplest example of a circular superstring loop stabilized by a chiral fermionic 
current. In Section 4, we show that, actually, the loops could have an
arbitrary shape and still be stabilized by the fermionic excitations on 
the worldsheet. We end with some conclusions and speculations on the 
relevance of these states.

\section{The superstring in the Green-Schwarz formalism}

In this section we will briefly review the Green-Schwarz formalism 
describing the dynamics of the superstring in ten dimensional flat 
spacetime. The basic idea of this formalism is to extend the usual Polyakov or Nambu-Goto
action written in terms of the bosonic target space coordinates of the string, 
$X^{\mu}$, to include ten dimensional fermionic degrees of freedom. These 
new degrees of freedom are two 32 component Majorana-Weyl spinors of 
$Spin(9,1)$, usually denoted by $\theta^{1A}$ and $\theta^{2A}$ 
($A = 1,\cdots, 32$), 
which have the same or opposite chirality defining type $IIB$ and $IIA$ 
superstrings respectively.  

{}The action contains two pieces: a kinetic term and a Wess-Zumino term. The 
kinetic term arises as the simplest generalization of the Polyakov action for 
the bosonic string by replacing $\partial_\alpha X^\mu$ by the combination
\begin{eqnarray}
 \Pi^{\mu}_{\alpha} = \partial_{\alpha} X^{\mu} - i \bar{\theta}^1 
\Gamma^{\mu} \partial_{\alpha} \theta^1- i \bar{\theta}^2 \Gamma^{\mu} 
\partial_{\alpha} \theta^2~,
\end{eqnarray}
that is invariant under global supersymmetry transformations
\begin{eqnarray}\label{su}
\delta \theta^{IA}=\epsilon^{IA}~,~~~~~\delta X^\mu=i\bar\epsilon^I\Gamma^\mu
\theta^I~, ~~~~~I=1,\ 2~.  
\end{eqnarray}
Thus, the kinetic term is given by
\begin{eqnarray}
{\cal{S}}_{Kin}&=& -{T\over {2\pi}} \int{d^2\sigma \sqrt{h} h^{\alpha \beta} 
 \Pi^{\mu}_\alpha \Pi_{\mu\beta}}~,
\end{eqnarray}
where $T$ is the string tension, $h^{\alpha \beta}$ is the two dimensional, 
supersymmetry inert, metric on
 the string worldsheet with $\alpha, \beta = 1,2 = \sigma, \tau$; 
$\mu$ denotes the spacetime coordinate index with range $0,\cdots, 9$, and 
$\Gamma^{\mu}$ are the ten dimensional $32\times 32$ Dirac matrices satisfying 
 $\{\Gamma^{\mu},\Gamma^{\nu}\} = 2 \eta^{\mu \nu}$. \footnote{A suitable real 
representation of the gamma matrices is given by:
\begin{eqnarray}
C=\Gamma^0=i\tau_2\otimes 1=\left(\begin{array}{cc} 0 & 1\\ 
-1 & 0 \end{array}\right)~, ~~~~
\Gamma^i=\tau_1\otimes\left(\begin{array}{cc} 0 & \gamma^i\\ 
\gamma^{iT} & 0 \end{array}
\right)~,~~~~
\Gamma^9= 
\tau_1\otimes\left(\begin{array}{cc} 1 & 0\\ 0 & -1 \end{array}\right)~,
\nonumber
\end{eqnarray}
where the $8\times8$ matrices $\gamma^i$ are given by:
\begin{eqnarray}\label{gammas}
\begin{array}{ll}\gamma^{1}=-i \tau_{2} \otimes  \tau_2  \otimes \tau_2~,&
~~~~~~\gamma^{5}= i \tau_{3} \otimes  \tau_2  \otimes 1~,\\
\gamma^{2}= i       1  \otimes  \tau_1  \otimes \tau_2~,&
~~~~~~\gamma^{6}= i \tau_{2} \otimes   1      \otimes \tau_1~,\\
\gamma^{3}= i       1  \otimes  \tau_3  \otimes \tau_2~,&
~~~~~~\gamma^{7}= i \tau_{2} \otimes   1      \otimes \tau_3~,\\
\gamma^{4}= i \tau_1   \otimes  \tau_2  \otimes  1    ~,&
~~~~~~\gamma^{8}=         1  \otimes   1      \otimes 1~,
\end{array}\nonumber
\end{eqnarray}
and $\tau_{1,2,3}$ are the Pauli matrices.}

On the other hand, this first term of the action by itself does not lead to a 
ten dimensional 
supersymmetric spectrum for the string, so one is forced to add the second, 
Wess-Zumino, term (also invariant under supersymmetry) that allows the 
possibility of finding a new symmetry of the total action. This is a local 
fermionic symmetry, Siegel 
($\kappa$) symmetry, that further restricts the degrees of freedom of the 
string. The total action then becomes,
\begin{eqnarray}\label{action}
S&=& S_{Kin} + S_{WZ}~,
\end{eqnarray}
where the Wess-Zumino term is given by
\begin{eqnarray}
S_{WZ}&=& {T\over{\pi}} \int d^2\sigma 
\left\{-i\epsilon^{\alpha \beta} \partial_{\alpha} X^{\mu} 
\left(\bar{\theta}^1\Gamma_{\mu} \partial_{\beta}\theta^1-
\bar{\theta}^2\Gamma_{\mu} \partial_{\beta}\theta^2\right)+ 
\epsilon^{\alpha \beta} \bar{\theta}^1 \Gamma^{\mu} \partial_{\alpha}\theta^1 
\bar{\theta}^2 \Gamma_{\mu}\partial_{\beta}\theta^2\right\}
\end{eqnarray}

{}The classical equations of motion for the fields
present in the action (\ref{action}), namely, $h_{\alpha \beta}, 
X^{\mu}, \theta^1$ and 
$\theta^2$ can be derived easily. Note, however, that in a general gauge, these 
equations would turn out
to be quite complicated as it also happens in the purely bosonic case 
if we do not choose the right gauge conditions. Just as for the
bosonic string \cite{GGRT}, it can be shown that the symmetries of
the action allow us to select a gauge (by fixing Weyl, 2D 
reparametrizations and residual semilocal conformal symmetries) in which
\begin{eqnarray}
X^+=X^t+ X^9= x^++l^2 p^+\tau ~,
\end{eqnarray}
where $l^2=1/T$.

{}On the other hand, we can also completely fix $\kappa$ symmetry by setting,
\begin{eqnarray}
&&\Gamma^+\theta^1=\Gamma^+\theta^2=0~,
\end{eqnarray}
where,
\begin{eqnarray}
\Gamma^+&=& \Gamma^0 + \Gamma^9~.
\end{eqnarray}
In this gauge (light-cone gauge) the equations of motion for the surviving 
dynamical degrees of freedom are extremely simple, since
they become,
\begin{eqnarray}
\left({{\partial^2}\over{\partial \sigma^2}} - {{\partial^2}\over{\partial 
\tau^2}}\right) X^i=0~,
\end{eqnarray}
\begin{eqnarray}
\left({{\partial}\over{\partial \tau}} + 
{{\partial}\over{\partial \sigma}}\right) S^{1a}=0~,
\end{eqnarray}
\begin{eqnarray}
\left({{\partial}\over{\partial \tau}} - 
{{\partial}\over{\partial \sigma}}\right) S^{2a}=0~,
\end{eqnarray}
where $X^i$ are the eight transverse degrees of freedom of the 
position of the string and $S^{1a}=\sqrt{p^+} \theta^1$ and 
$S^{2a}=\sqrt{p^+} \theta^2$, with $a= 1,\cdots, 8$, are the left and 
right-moving 
Majorana-Weyl 
fermions in the ${\bf 8_s}$ and ${\bf 8_s}$ (or ${\bf 8_c}$) $Spin(8)$ spinor 
representation respectively in type IIB (or A).

We can therefore write the general solution for a closed superstring
loop using the standard decomposition into left and right-movers 
\begin{eqnarray}
X^j&=&X^j_R+X^j_L~,\\
X^j_R&=&{1\over 2}x^j+{1\over 2} l^2 p^j(\tau-\sigma)+
{i\over 2}l
\sum_{n\not = 0} {1\over n} \alpha_n^j{\rm e} ^{-2in(\tau-\sigma)}~,\\
X^j_L&=&{1\over 2}x^j+{1\over 2} l^2 p^j(\tau+\sigma)+{i\over 2}l
\sum_{n\not = 0} {1\over n} \tilde\alpha_n^j{\rm e}^{-2in(\tau+\sigma)}~,
\end{eqnarray}
where $j=1, \cdots, 8$, and $x^j$ and $p^j$ are the center of mass
position and momentum of the loop. Similarly we can decompose the 
fermionic coordinates as,
\begin{eqnarray}
S^{1a}&=&\sum_{n=-\infty}^{\infty} S^a_n {\rm e} ^{-2in(\tau-\sigma)}~,
\nonumber\\
S^{2a}&=&\sum_{n=-\infty}^{\infty}\tilde S^a_n {\rm e} ^{-2in(\tau+\sigma)}.
\end{eqnarray}
where the coefficients satisfy anti-commutation relations, namely,
\begin{eqnarray}\label{ar}
\{S^a_{m},S^b_n\}=\delta^{ab}\delta_{m+n,0}
\end{eqnarray}
This is, of course, not the whole story, since we should also have
an equation for the metric $h_{\alpha \beta}$ coming from the general
action. Once we restrict this equation to our particular gauge choice, 
namely the light-cone gauge, we obtain the following constraint equations,
\begin{eqnarray}
\label{Virasoro-constraints}
\alpha^{-}_{n}&=&{1\over 2 l p^{+}}
\sum_{m=Z}\left(\alpha^i_{n-m}\alpha^i_m+\left(m-{n\over 2}\right)
 S^a_{n-m} S^a_m \right)\label{vir}\\
\tilde \alpha^{-}_{n}&=&{1\over 2 l p^{+}}
\sum_{m=Z}\left(\tilde \alpha^i_{n-m}\tilde \alpha^i_m+
\left(m-{n\over 2}\right)
\tilde S^a_{n-m}\tilde S^a_m\right) \label{tvir}
\end{eqnarray}
where $\alpha^{-}_{n}$ and $\tilde \alpha^{-}_{n}$ are the expansion 
coefficients
of the light-cone coordinate, $X^{-}$, namely,
\begin{eqnarray}
X^{-}&=&x^- + l^2 p^- \tau+
{i\over 2}l
\sum_{n\not = 0} {1\over n} 
\left(\alpha_n^-{\rm e} ^{-2in(\tau-\sigma)}+ 
\tilde\alpha_n^-{\rm e}^{-2in(\tau+\sigma)}\right)
\end{eqnarray}
Finally consistency of these equations requires that,
\begin{eqnarray}\label{H}
\left(-l^2 p^+ p^- +l^2(p^i)^2+8N_{L,R}\right)=0~,
\end{eqnarray}
for both left and right movers,
\begin{eqnarray}
N_L&=&\sum_{n=1}\alpha_{-n}^i\alpha_{n}^i+nS_{-n}^aS_n^a~,\nonumber\\
N_R&=&\sum_{n=1}\tilde\alpha_{-n}^i\tilde\alpha_{n}^i+n\tilde S_{-n}^a\tilde 
S_n^a~.
\end{eqnarray}
{}In the remainder of the paper we work at the quantum level, therefore these 
constraints should be satisfied on the quantum states.

\section{A superstring loop}

We will consider a quantum state $|\phi\rangle$ of the Green-Schwarz 
superstring in the light-cone gauge such that it has no $X^9$ excitations, 
in other words we will have that $p^+=p^-=p$ which in turn means that,
\begin{eqnarray}\label{lc}
X^+=X^-=l^2 p\tau~.
\end{eqnarray}
As we explained before in the light-cone gauge, the string is built 
by acting with several bosonic and fermionic operators on the right 
and left moving vacuum sectors. In our case, we are interested in
studying the state with bosonic excitations only in the left moving
sector and fermionic excitations present only in the right moving
sector. In particular we can discuss the simplest such a state
which we construct in the following way,
\begin{eqnarray}\label{state}
|\phi\rangle={\rm e}^{v\alpha_{-1}^x-v^*\alpha_1^x}
{\rm e}^{w\alpha_{-1}^y-w^*\alpha_1^y}|0\rangle_L\otimes 
V^1_{h+1} V^1_{h+2}\cdots V^1_{0} |0\rangle_R~,
\end{eqnarray}
where $v$ and $w$ are complex parameters and $h$ is a 
negative integer and where we have defined
\begin{eqnarray}
V^q_n&=&\tilde S^{2q-1}_n+i\tilde S^{2q}_n~,\nonumber\\
\bar V^q_n&=&\tilde S^{2q-1}_n-i\tilde S^{2q}_n~,
\end{eqnarray}
with $q=1,\cdots, 4$.

The right-moving part of the state is reminiscent of a $q-$vacuum as 
constructed in \cite{fms} for a general $b-c$ system. Starting from a 
vacuum
\footnote{The vacuum states of the GS superstring in each sector 
($R$ and $L$)
are a spanned by a set of eight bosonic states $|i\rangle$ in the 
vector representation of $Spin(8)$ and eight fermions $|a\rangle$ in the 
{\bf $8_s$} (or {\bf $8_c$}) representation. In order to build
the state 
$|\phi\rangle$ we can choose any of these as our vacuum.}
\begin{eqnarray}
\tilde S^a_n|0\rangle_R=0~, ~~~n>0~,
\end{eqnarray} 
we can build this part of the state as in (\ref{state}),
\begin{eqnarray}
|h\rangle_R= V^1_{h+1} V^1_{h+2}\cdots V^1_{0} |0\rangle_R~,
\end{eqnarray}
which has the property that for negative $h$ it satisfies
\begin{eqnarray}
V^1_n|h\rangle_R&=&0~, ~~~n>h~,\nonumber\\
\bar V^1_n|h\rangle_R&=&0~, ~~~n\geq -h~.
\end{eqnarray} 
Using (\ref{ar}) we can also show that the new operators $V$ and $\bar V$ 
satisfy the following anticommutation relations, 
\begin{eqnarray}
\{\bar V^p_n, V^q_m\}= 2 \delta_{m+n,0}\delta^{pq}~.
\end{eqnarray}

We will now show that it is indeed possible for this 
state to fulfill all the physical constraints described in 
the previous section by fixing the parameters in the state 
construction, namely, $v,w,h$.

From the constraint (\ref{H}) for the left-movers we obtain the condition
\begin{equation}
\label{constraint1}
-l^2p^2+8\left( |v|^2+|w|^2\right)=0~.
\end{equation}
Taking into account that,
\begin{eqnarray}
\sum_{n=1} n\tilde S_{-n}^a\tilde S_n^a|h\rangle_R={{h(h+1)}\over{2}}
|h\rangle_R~,
\end{eqnarray}
we see that the Hamiltonian constraint (\ref{H}) for the right-movers 
imposes the following relation,
\begin{equation}\label{Hsol}
-l^2p^2+4{h(h+1)} =0~,
\end{equation}
Also, consistency of the Virasoro condition (\ref{vir}) with eq. (\ref{lc})
implies (only $n=2$ gives a nontrivial equation):
\begin{equation}
\label{constraint2}
v^2+w^2=0~.
\end{equation}
Finally, it is not difficult to check that eq. (\ref{tvir}) 
does not give any other condition for our state. Choosing, therefore, 
the parameters $v,w,h$ fulfilling eqns. (\ref{constraint1}) (\ref{Hsol}) 
and (\ref{constraint2}) would completely fix the state we are looking for.

Having done that, the position of the string in this state is given by:
\begin{eqnarray}
\left< X^t\right>&=&l^2p\tau\\
\left< X^x\right>&=&{i\over 2}l\langle\phi|
-\alpha_{-1}^x{\rm e}^{2i(\tau-\sigma)}+
 \alpha_{1}^x{\rm e}^{-2i(\tau-\sigma)}
|\phi\rangle\nonumber\\
&=&{i\over 2}l \left(-v^*{\rm e}^{2i(\tau-\sigma)}+
v{\rm e}^{-2i(\tau-\sigma)}\right)\\
\left< X^y\right>&=&{i\over 2}l\langle\phi|
-\alpha_{-1}^y{\rm e}^{2i(\tau-\sigma)}+
 \alpha_{1}^y{\rm e}^{-2i(\tau-\sigma)}
|\phi\rangle\nonumber\\
&=&{i\over 2}l \left(-w^*{\rm e}^{2i(\tau-\sigma)}+
w{\rm e}^{-2i(\tau-\sigma)}\right).
\end{eqnarray}

{}Choosing real $lp=4v=2\sqrt{h(h+1)}$ and $w=-iv$ we satisfy the 
constraints and the string has spacetime parametrization given by
\begin{eqnarray} 
\left<X^\mu\right>&=&(4 l v\tau, l v~ {\rm sin}~2(\tau-\sigma),l v~ 
{\rm cos}~2(\tau-\sigma),0,\dots)~.
\end{eqnarray}
A fixed circle of radius $l v$, which for large enough values of the parameter 
$v$ can be of cosmological size. We will address the issue of the stability of 
this loop in section 5. 

\section{Loops of arbitrary shape}

The loop considered on the previous section had a circular shape but 
we will now show that this can be easily generalized to states with arbitrary 
shape.
Let us consider a loop constructed from a general coherent state of
the form:
\begin{eqnarray}
\label{general-state}
|\phi\rangle={\rm e}^{\cal{A}}|0\rangle_L\otimes |h\rangle_R~,
\end{eqnarray}
where 
\begin{eqnarray}
\cal{A}&=&\sum_{n=1}^N C^i_n~\alpha^i_{-n}-{\rm h.~c.}~,
\end{eqnarray}
and $C^i_{n}$ are complex numbers satisfying,
\begin{eqnarray}
C^i_n&=& \left(C^i_{-n}\right)^*~.
\end{eqnarray}
In order for this state to parametrize a static loop in
spacetime we should impose the constraint
equations described by (\ref{Virasoro-constraints}), which in our state 
reduce to,
\begin{eqnarray}
\alpha^{-}_{n}|\phi\rangle&=&{1\over 2 l p^{+}}
\sum_{m=Z}\alpha^i_{n-m}\alpha^i_m|\phi\rangle=0~.
\end{eqnarray}
This in turn means that the coefficients $C^i_{n}$
should obey the following relations,
\begin{eqnarray}
\sum_{m=n-N}^N  C^i_{n-m}C^i_{m}  &=&0~,
\end{eqnarray}
for $n=1,\cdots 2N$. Similarly, we should impose the Hamiltonian constraint to 
this state which in this case becomes,
\begin{eqnarray}
\sum_{m=n-N}^N  C^i_{-m}C^i_{m}  &=&const~.
\end{eqnarray}
In order to see what restrictions these constraints have
on its shape, we calculate the expectation value of the position of the
string at a fixed time, $X^j(\sigma, \tau=0)$,
\begin{eqnarray}
\left< X^j\right>&=&l~\sum^{N}_{n=-N} {1\over{2in}}~C_n^j~{\rm e}^{2in\sigma}~.
\end{eqnarray}
This shows that the constraints on the coefficients $C_n^i$ are just 
imposing that $\sigma$ parametrizes the position of the string
such that 
\begin{eqnarray}
\label{parametrization-constraint}
\left|{{d\left< X^j\right>}\over {d\sigma}}\right|&=& const~.
\end{eqnarray}
Since for any given function we can always find a parametrization
of the position of the string that satisfies 
(\ref{parametrization-constraint}), we conclude that we 
will always be able to describe such a string with a coherent state 
like the one presented in (\ref{general-state}).

\section{Concluding remarks}

We have shown that there are states of the perturbative superstring 
spectrum which can be identified as string loops stabilized
by the presence of fermionic excitations on the string 
worldsheet. The energy momentum tensor associated with the
fermions creates a mechanical backreaction that allows these strings 
to remain static in spacetime. This is exactly the same effect that occurs in field
theory models of cosmic strings where the fermionic degrees of 
freedom on the string worldsheet come from zero modes trapped on the 
vortex solution \cite{Witten}.

On the other hand, field theory cosmic strings can also have bosonic 
zero modes associated with the phase of a condensate living 
on the string. This degree of freedom can form neutral superconducting loops,
the so called vortons \cite{Davis}. It is therefore natural to wonder whether 
there are similar configurations in string theory.

In this case, the situation is a little bit more complicated, since there are
no bosonic condensates coupled to the superstring. It is nevertheless still 
possible to find stable string configurations by considering the 
effect on the loop motion of the excitations of the position of the string along 
some perpendicular direction. These excitations can be thought of as a bosonic current on the
loop worldsheet due to the Goldstone bosons associated with the breaking of translational 
invariance along those directions, either along a plane perpendicular to the string in 
ten dimensional spacetime \cite{CIR} or along some compactified direction \cite{BPI-1}.

On the other hand, we have also shown that the states we discuss in this
paper are characterized by their fermionic content and that the string shape
in space is totally arbitrary in the plane perpendicular to $X^9$. This is another 
characteristic shared with their classical superconducting string counterparts \cite{chiral1, chiral2}, 
a reflection of the fact that we are basically dealing with the same physical effect.

The existence of these arbitrary shaped loops of fundamental strings could have 
important cosmological consequences in the recently discussed models of brane
inflation, where loops of fundamental strings are expected to be produced at
cosmological scales \cite{Gia-Alex-2, CMP}. Some of these loops will be produced with some
fermionic excitations and could end up as the states we have been discussing.
We note, in this regard, that we certainly expect large loops to be very stable
since they preserve some fraction of supersymmetry in the straight string limit \footnote{ We want to thank
Gia Dvali and Michele Redi for useful discussions on this point.} 

One can see this by looking at the supersymmetry transformations (\ref{su}), which
in the lightcone gauge split in two classes. Those with $\epsilon$ parameter 
such that the gauge choice is preserved, which give rise to eight supersymmetries 
in each sector ($L$ and $R$), of the form
\begin{eqnarray}\label{pres}
\delta S^a=\sqrt{2p^+}\eta^a~,~~~~~\delta X^i=0~,
\end{eqnarray}
and those that need a compensating $\kappa$ transformation which are of the 
form
\begin{eqnarray}\label{comp}
\delta S^a=-i\rho\cdot \partial X^i\gamma^i_{a\dot a}\epsilon^{\dot a}
\sqrt{2p^+}~,~~~~~\delta X^i=\sqrt{2\over p^+}\gamma^i_{a \dot a}\bar
\epsilon^{\dot a}S^a~.
\end{eqnarray}
One immediately notices that in the absence of bosonic excitations in the $i$
directions ({\em i.e.}, an infinite straight string), one is free to put 
right-moving fermionic excitations and still preserve the 8 supersymmetries 
(\ref{comp}) on the left-moving sector. These are 1/4 BPS states which support 
fermionic excitations. Thus, one would expect that at least
a string loop of huge size and large curvature radius should be long-lived.
This has also been realized recently in the context of fermionic 
zero modes on supersymmetric cosmic strings in \cite{gia}.

Finally, we focused in this paper in states of type $IIA$ and $IIB$ superstrings. But 
the generalization to other models is expected. The states we discussed will 
also be present in Heterotic strings (just taking the fermions in the 
supersymmetric side), and a generalization to $D1$ brane states should not be 
difficult to achieve. Also, this construction (neglecting issues of 
quantization) should hold in spacetimes with dimensionality 3, 4 and 6 
where a Green-Schwarz superstring action can be constructed.

\section*{Acknowledgments}

{} We would like to thank Gia Dvali, Jaume Garriga, Marta Gomez-Reino, Martin 
Kruczenski, David Mateos, Ken Olum, Joseph Polchinski, Massimo Porrati, 
Michele Redi and Alexander Vilenkin for many useful discussions. 
A.I. is supported by funds provided by New York 
University and J.J.B.-P. is supported by the James Arthur Fellowship at NYU.

\end{document}